\documentclass[twocolumn]{aastex6}

\shorttitle{On the Orbital Inclination of Proxima Centauri
  \lowercase{b}}
\shortauthors{Stephen R. Kane et al.}
\slugcomment{Submitted for publication in the Astrophysical Journal
  Letters}

\begin{document}

\title{On the Orbital Inclination of Proxima Centauri \lowercase{b}}

\author{Stephen R. Kane}
\affil{Department of Physics \& Astronomy, San Francisco State
  University, 1600 Holloway Avenue, San Francisco, CA 94132, USA}
\email{skane@sfsu.edu}

\author{Dawn M. Gelino}
\affil{NASA Exoplanet Science Institute, Caltech, MS 100-22, 770 South
  Wilson Avenue, Pasadena, CA 91125, USA}

\author{Margaret C. Turnbull}
\affil{SETI Institute, Carl Sagan Center for the Study of Life in the
  Universe, Off-Site: 2801 Shefford Drive, Madison, WI 53719, USA}


\begin{abstract}

The field of exoplanetary science has seen discovery rates increase
dramatically over recent years, due largely to the data from the {\it
  Kepler} mission. Even so, individual discoveries of planets orbiting
nearby stars are very important for studies of characterization and
near-term follow-up prospects. The recent discovery of a terrestrial
planet candidate orbiting Proxima Centauri presents numerous
opportunities for studying a Super-Earth within our own stellar
backyard. One of the remaining ambiguities of the discovery is the
true mass of the planet since the discovery signature was obtained via
radial velocities. Here we describe the effect of orbital inclination
on the Proxima Centauri planet, in terms of mass, radius, atmosphere,
and albedo. We calculate the astrometric, angular separation, and
reflected light properties of the planet including the effects of
orbital eccentricity. We further provide dynamical simulations that
show how the presence of additional terrestrial planets within the
Habitable Zone varies as a function of inclination. Finally, we
discuss these effects in the context of future space-based photometry
and imaging missions that could potentially detect the planetary
signature and resolve the inclination and mass ambiguity of the
planet.

\end{abstract}



\section{Introduction}
\label{intro}

Exoplanet discoveries have increasingly focused on terrestrial
planets as detection capabilities continue to improve. For example,
the planet yield from the {\it Kepler} mission that are of primary
interest are those terrestrial planets that lie in the Habitable Zone
(HZ) of their host stars \citep{kan16}. For non-transiting planets,
the radial velocity (RV) method continues to be the primary method to
detect terrestrial planets suitable for follow-up
characterization. For example, the star HD~40307 harbors a system of
super-Earths discovered by the RV technique \citep{may09}, one of
which is known to lie within the HZ of the star \citep{tuo13}.

The closest exoplanet to the Solar System was recently identified by
\citet{ang16}, orbiting the closest star, Proxima Centauri. Proxima is
a late-type flare star with a rotation period of $\sim$84 days
confirmed photometrically \citep{ang16} and spectroscopically
\citep{coll16,rob16}. The associated planet was detected through a
long-term RV campaign and found to have an orbital period of 11.186
days, a semi-major axis of 0.0485~AU, and a minimum mass $\sim$30\%
larger than the Earth. Formation scenarios for the planet include
possible perturbations from close encounters with the Alpha Centauri
stellar components as a possible explanation for the relatively high
planetary orbital eccentricity \citep{bar16,cole16}. The size of the
planet remains unknown since transits have been effectively ruled out
\citep{ang16,dav16,kip16}. However, even though the inclination, true
mass, and radius are unknown, the planet is likely terrestrial. This
has led to the exploration of potential habitability conditions and
detectable biosignatures \citep{bar16,mea16,rib16,tur16}, including
the prospect of life in high UV environments \citep{oma16}.

Here we present an investigation of the effects of the inclination of
the Proxima Centauri b orbital plane relative to the line of
sight. The effects of the inclination on the mass of the planet and
related physical properties are described in Section~\ref{effect}. The
astrometric signature of the planet as a function of orbital
inclination is considered in Section~\ref{ast}. In Section~\ref{sep},
we provide calculations of the star--planet angular separation as a
function of inclination and orbital phase. Section~\ref{ref} discusses
the dependence of inclination on the expected phase varaitions due to
reflected light and related effects. Section~\ref{stab} presents the
results of a dynamical simulation that constrains the presence of
other potential terrestrial planets within the HZ of the host star. In
Section~\ref{obs} we discuss observable imaging signatures of the
planet and mission requirements to achieve a detection.


\section{The Effect of Inclination on Planetary Properties}
\label{effect}

The minimum mass of the Proxmina planet measured from the RV work of
\cite{ang16} is $M_p \sin i = 1.27$ Earth masses. The range of masses
and radii for which a planet can reasonably be expected to be
terrestrial, has been studied in detail, thanks largely to the planet
yield from the {\it Kepler} mission \citep{wei14,dre15b,rog15}. Many
of these studies find that there is evidence of a density transition
that occurs $\sim$1.5--2.0 Earth radii ($R_\oplus$) whereby the
composition of objects larger than this become dominated by volatile
rather than refractory materials. Using the mass-radius relationship
of \citet{wei14}, we estimate that this transition corresponds to
$\sim$3.9--5.1 Earth masses ($M_\oplus$). In order for the mass to
exceed this range, the orbital inclination would need to satify $i <
14.4\degr$. Assuming randomly oriented orbits, and excluding the 1.5\%
transit probability \citep{ang16}, the probability that the planet
lies in the terrestrial regime is $\sim$84\%.

Apart from increasing the radius of the planet, decreasing the orbital
inclination and thus increasing the planetary mass also has an effect
on the atmospheric properties (see \citet{mad16} and references
therein). For a given insolation flux, the atmospheric composition
will determine the resulting chemistry and the formation of reflective
layers in the upper atmospheric layers. Of particular relevance is the
transition from terrestrial to giant planet whereby the dominant
atmospheric components change from heavy molecules (H$_2$O, CO$_2$,
N$_2$) to high H/He abundances. The impact of these on observations
lies primarily in the affect on the resulting albedo and contrast
ratios of the planet to the host star. These factors are discussed in
more detail in Sections~\ref{ref} and \ref{obs}.


\section{Astrometric Signature}
\label{ast}

\begin{figure}
  \includegraphics[angle=270,width=8.5cm]{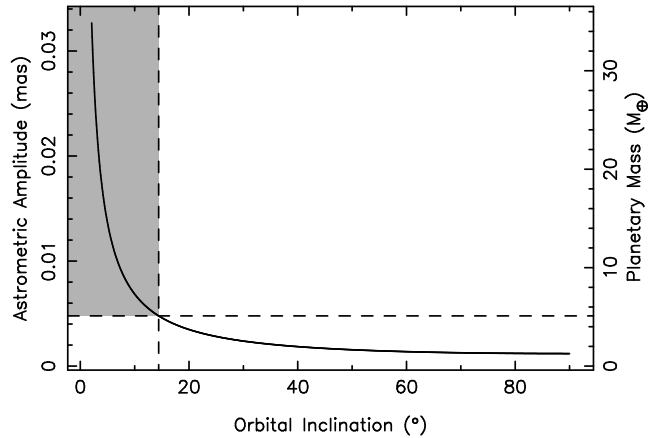}
  \caption{The astrometric amplitude of Proxima Centauri due to the
    orbit of the planet as a function of orbital inclination. The
    corresponding planetary mass is shown on the right-side y-axis of
    the plot. The dashed cross-hairs indicate the location where the
    mass is at the terrestrial/gas planet threshold (see
    Section~\ref{effect}). The gray shaded region is then where an
    astrometric detection would confirm that the planet is too massive
    to be considered terrestrial.}
  \label{astfig}
\end{figure}

A change in planetary mass has implications for the expected
astrometric signature. The amplitude of an astrometric signature is
given by
\begin{equation}
  \alpha = \left( \frac{M_p}{M_\star} \right) \left(
  \frac{a}{\mathrm{1 AU}} \right) \left( \frac{d}{\mathrm{1 pc}}
    \right)^{-1} \ \mathrm{arcsec}
\end{equation}
where $M_p$ and $M_\star$ are the planetary and stellar masses
respectively, $a$ is the semi-major axis in AUs, and $d$ is the
distance to the system. In Figure~\ref{astfig} we plot the astrometric
signature of the Proxima planet as a function of the orbital
inclination. We include the corresponding mass of the planet on the
right-hand y-axis of the plot, and the location of the terrestrial/gas
planet threshold (dashed lines), as discussed in
Section~\ref{effect}. The gray region of the plot thus highlights the
region where an astrometric detection of that magnitude would resolve
the $\sin i$ ambiguity in favor of the planet being non-terrestrial in
nature.

\begin{figure*}
  \begin{center}
    \begin{tabular}{cc}
      \includegraphics[angle=270,width=8.5cm]{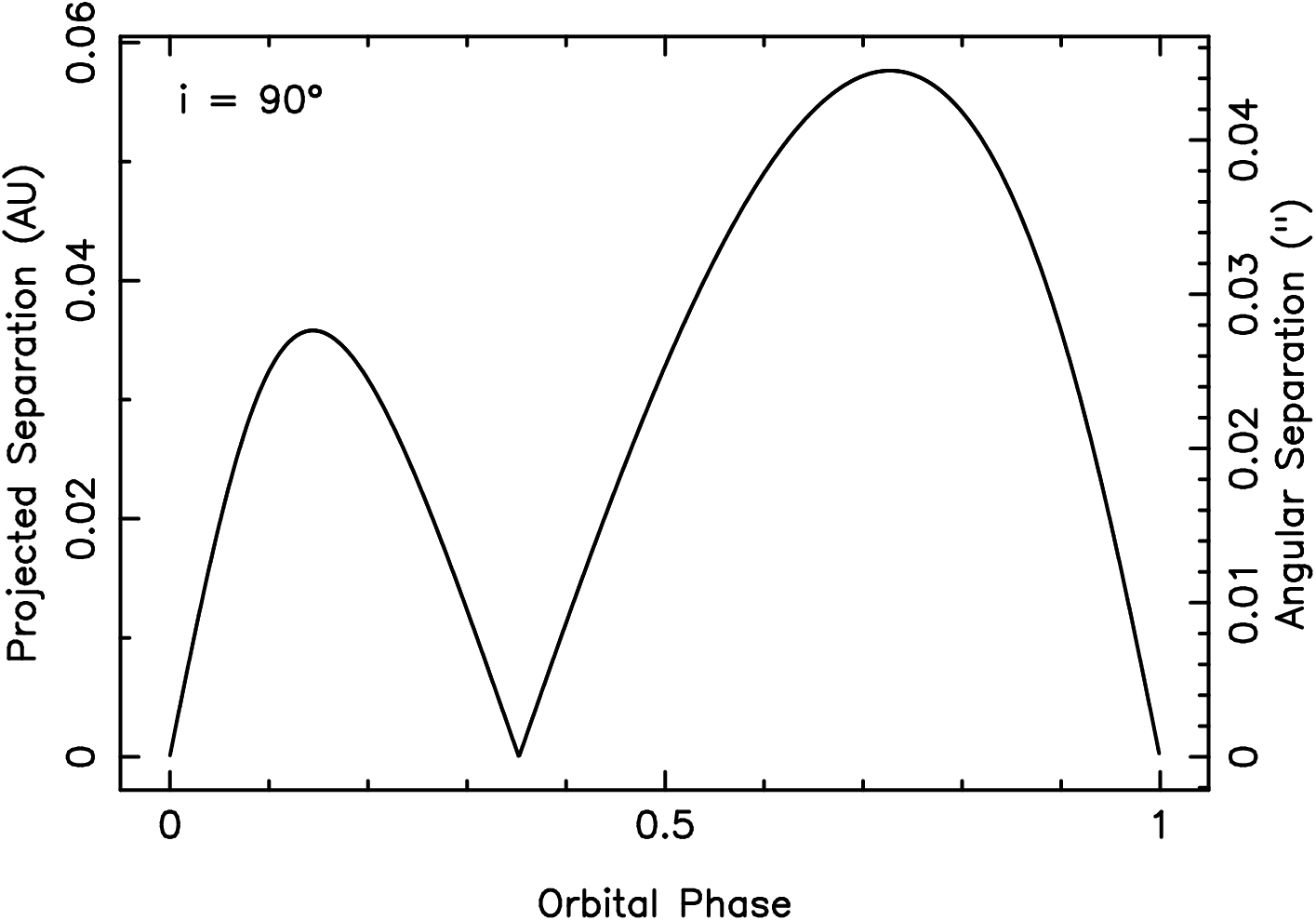} &
      \includegraphics[angle=270,width=8.5cm]{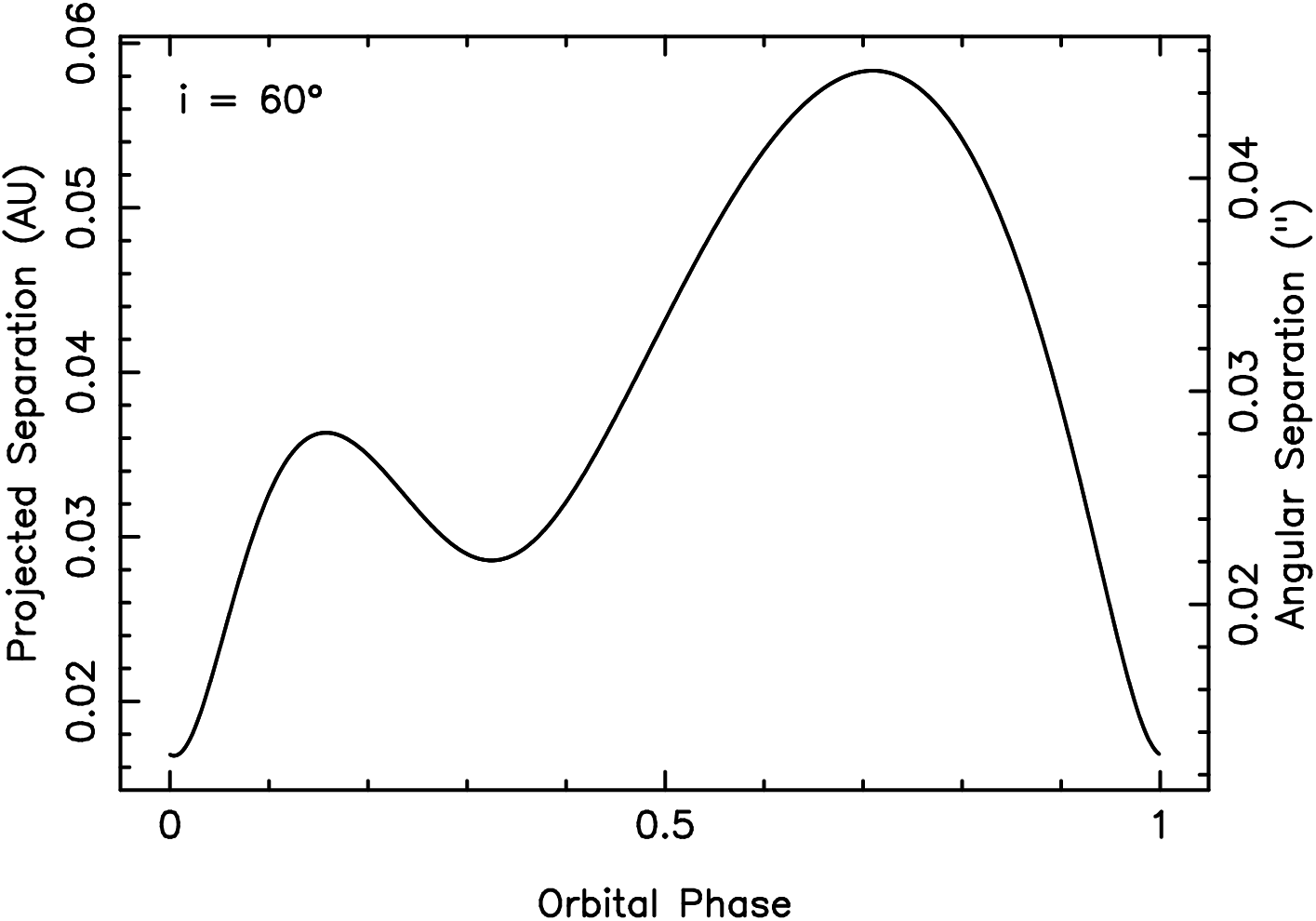} \\
      \includegraphics[angle=270,width=8.5cm]{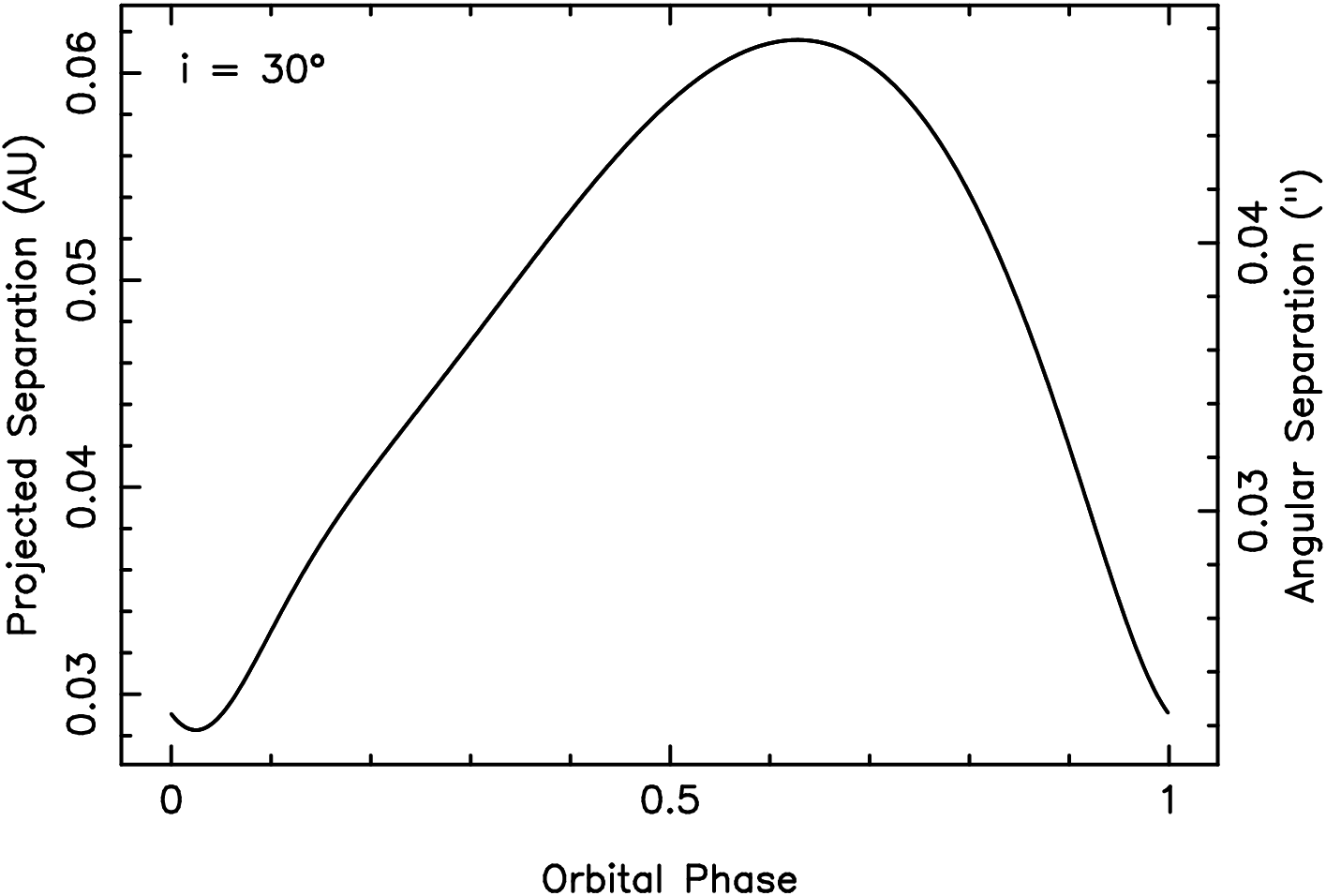} &
      \includegraphics[angle=270,width=8.5cm]{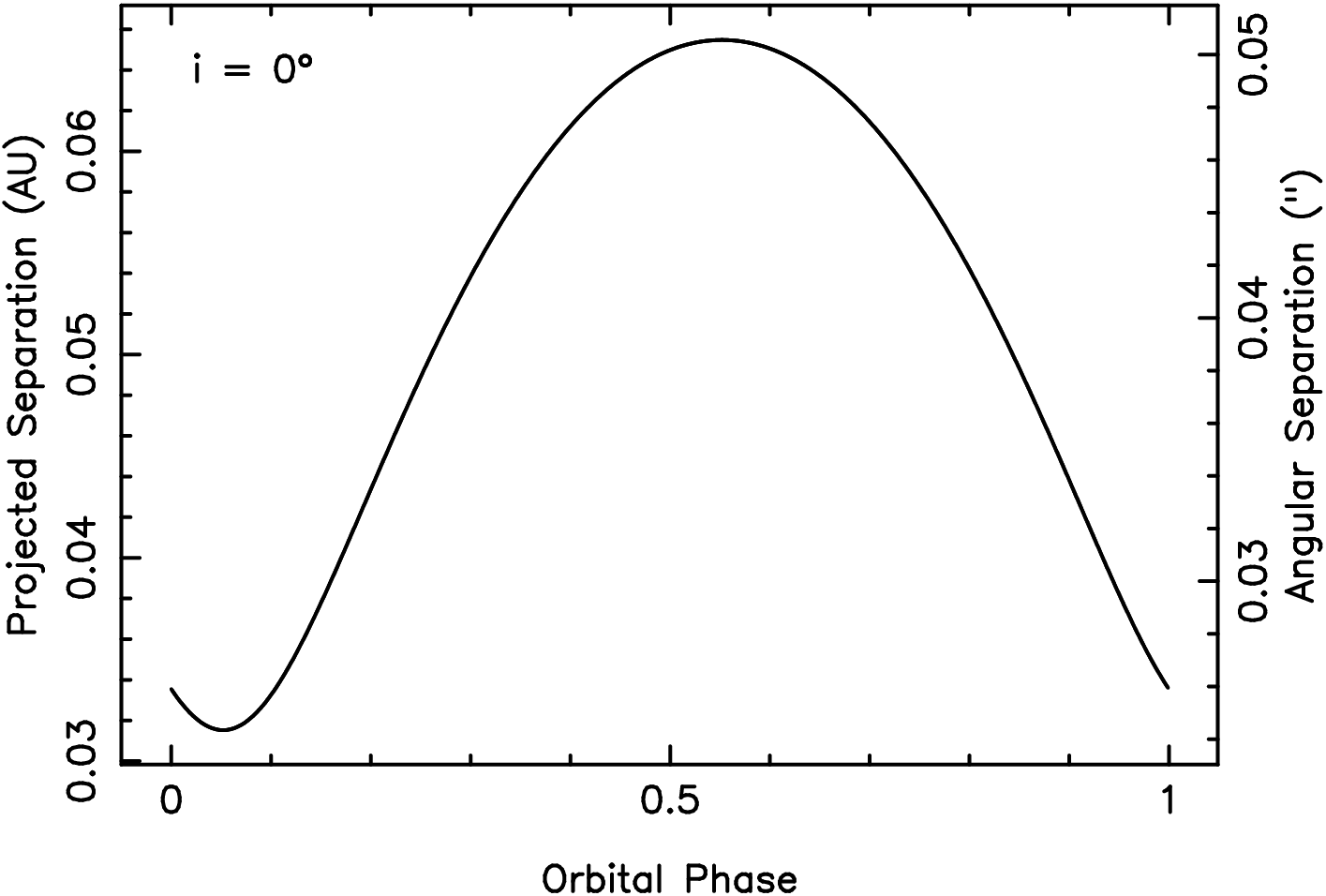}
    \end{tabular}
  \end{center}
  \caption{Projected and angular separation of Proxima Centauri b from
    the host star, assuming inclinations of $i = 90\degr$ (top-left),
    $i = 60\degr$ (top-right), $i = 30\degr$ (bottom-left), and $i =
    0\degr$ (bottom-right). An orbital phase of zero corresponds to
    the location of superior conjunction. These separations need to be
    taken into account when planning future observations.}
  \label{sepfig}
\end{figure*}

The {\it Gaia} mission \citep{pru16} is currently in the process of
astrometric data releases \citep{bro16,lin16}\footnote{\tt
  http://gea.esac.esa.int/archive}. These data will undoubtedly
contribute enormously to exoplanet science and the exoplanet detection
capabilities of {\it Gaia} have been previously investigated by
\citet{per14}. The expected science performance of {\it Gaia} has been
described by \citet{deb12} and is also available at the European Space
Agency (ESA) web site for the mission\footnote{\tt
  http://www.cosmos.esa.int/web/gaia/science-performance}. From these
sources, the expected astrometric precision for bright ($5 < V < 14$)
M dwarf stars is 5--16~$\mu$as. This is more than sufficient to
adequately sample the gray region of Figure~\ref{astfig} and perhaps
detect the planetary signature within the terrestrial regime. A
limitation of such analysis is the relatively short orbital period of
the planet in comparison to the cadence of the {\it Gaia}
observations. However, combining the astrometry with further RVs will
be able to resolve the full orbital solution for the planet
\citep{tuo09}.


\section{Angular Star--Planet Separation}
\label{sep}

The Proxima Centauri system is likely to be an attractive target for
planned imaging missions and the angular separation of the planet from
the host star will be a key part of those observations. The planning
of those observations is particularly important if indeed the
eccentricity of the planetary orbit is close to upper limit of $e =
0.35$ found by \citet{ang16}, since even face-on orbits will have a
time-dependence to the star--planet separation. It is also worth
noting that there is a bias toward higher orbital eccentricities in RV
exoplanet surveys \citep{zak11}, thus increasing the likelihood of an
eccentricity for Proxima Cenaturi b that lies closer to the maximum
value. Using the methodology of \citet{kan13}, we calculate the
angular separation of the planet over one complete orbit. Shown in
Figure~\ref{sepfig} are the projected and angular separations of the
planet from the host star for four possible orbital inclinations,
including edge-on ($i = 90\degr$) and face-on ($i = 0\degr$) viewing
angles. An orbital phase of zero corresponds to the location of
superior conjunction. These calculations assume both the eccentricity
of $e = 0.35$ and the argument of periastron of $\omega = 310\degr$
given by \citet{ang16}.

As expected, the maximum angular star--planet separation
($\sim$65~mas) occurs for the case of $i = 0\degr$. However, this
separation is only slightly larger than those of other
inclinations. The primary consideration for the different orbital
inclinations are the timing of the observations, which can result in
negligible star--planet separations, particularly for $i >
60\degr$. This is discussed in the context of future missions in
Section~\ref{obs}.


\begin{figure*}
  \begin{center}
    \begin{tabular}{cc}
      \includegraphics[angle=270,width=8.5cm]{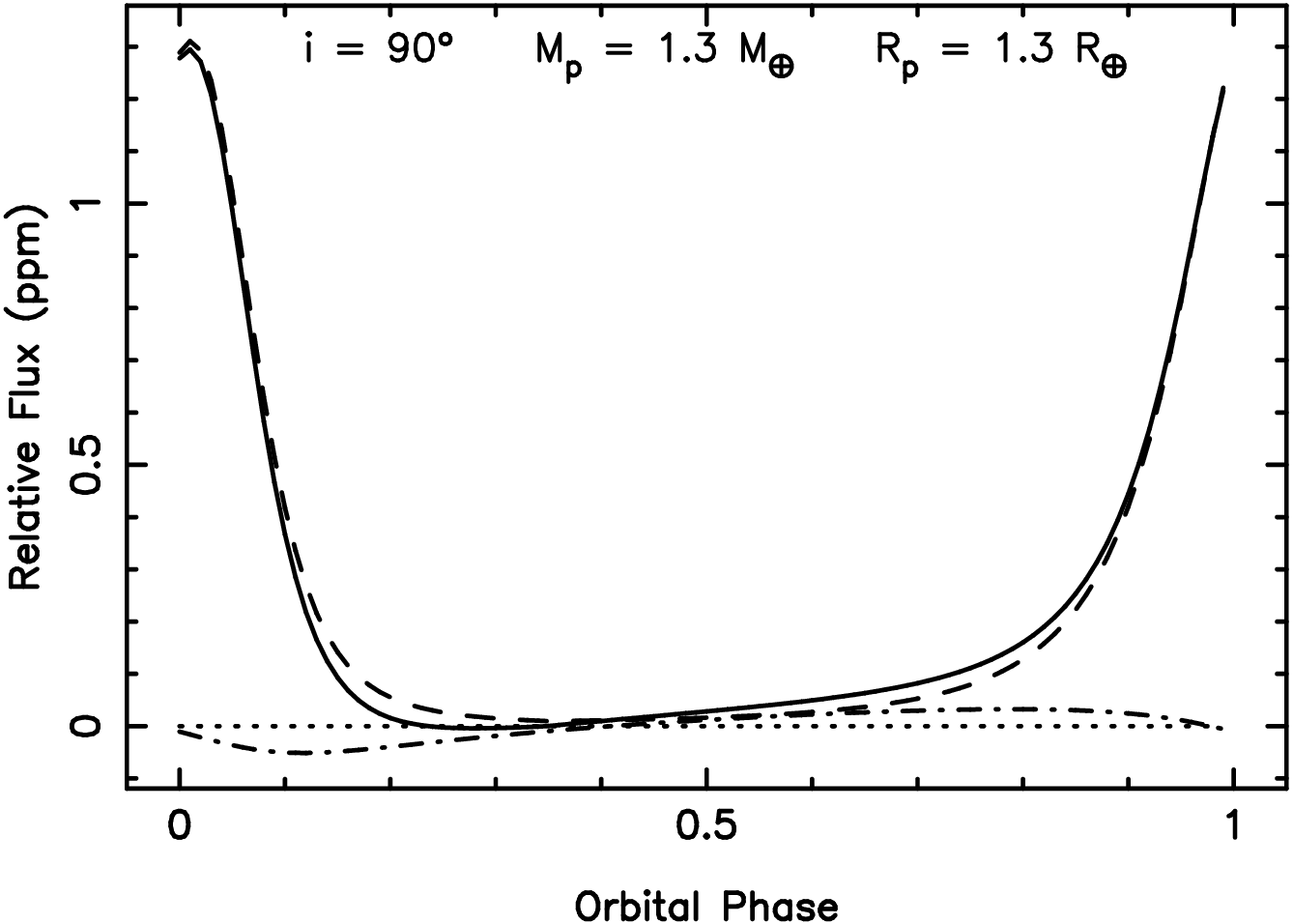} &
      \includegraphics[angle=270,width=8.5cm]{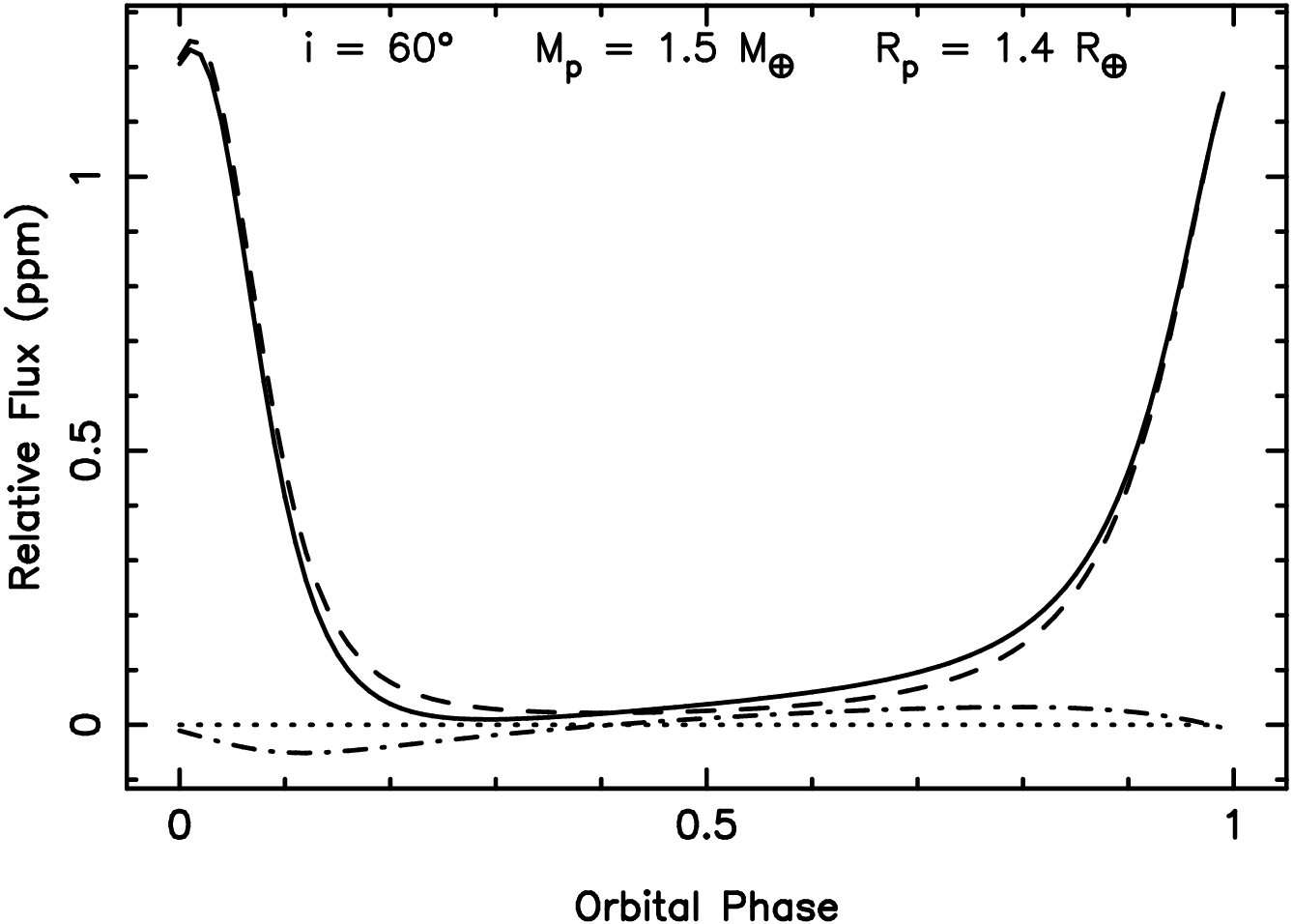} \\
      \includegraphics[angle=270,width=8.5cm]{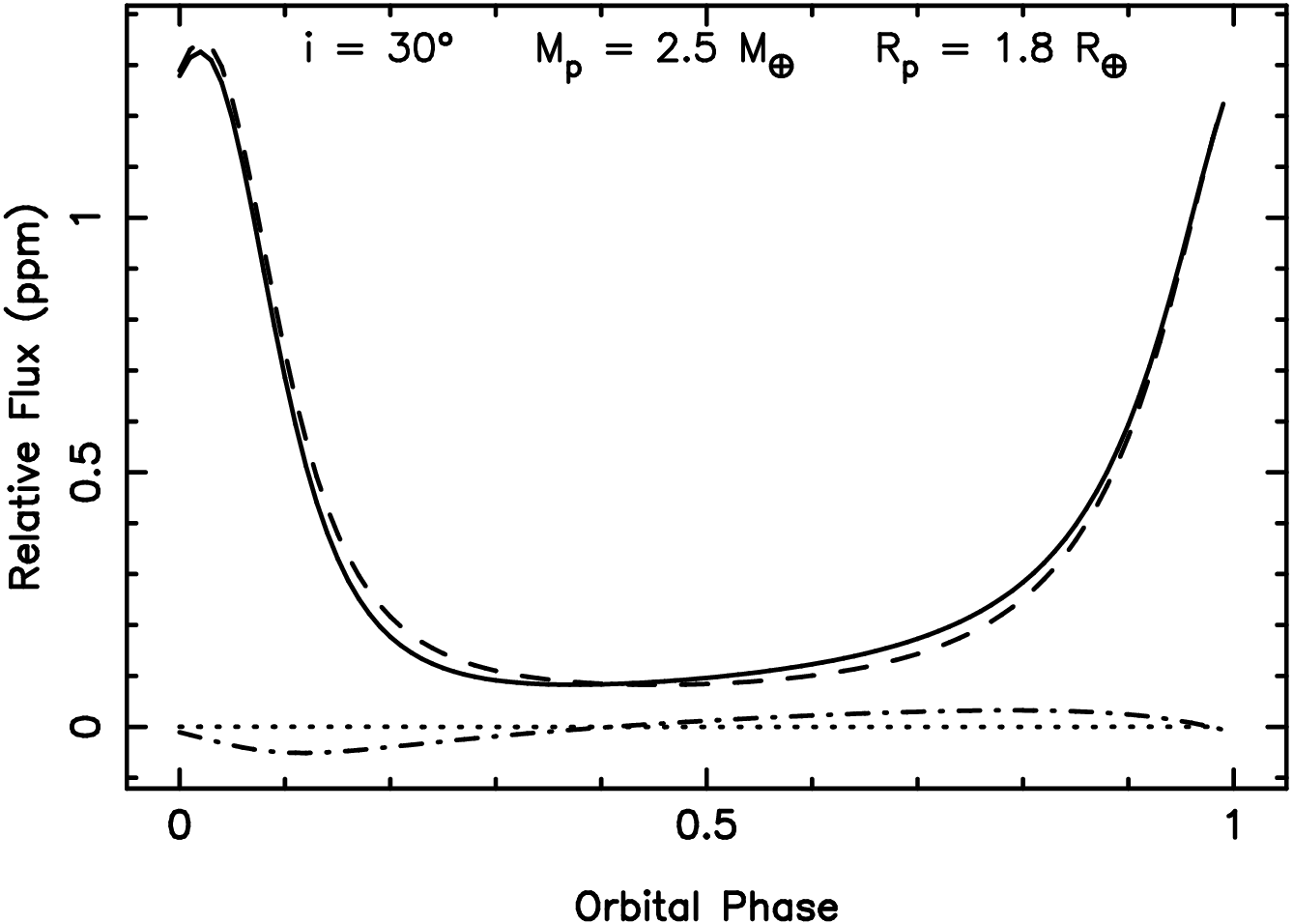} &
      \includegraphics[angle=270,width=8.5cm]{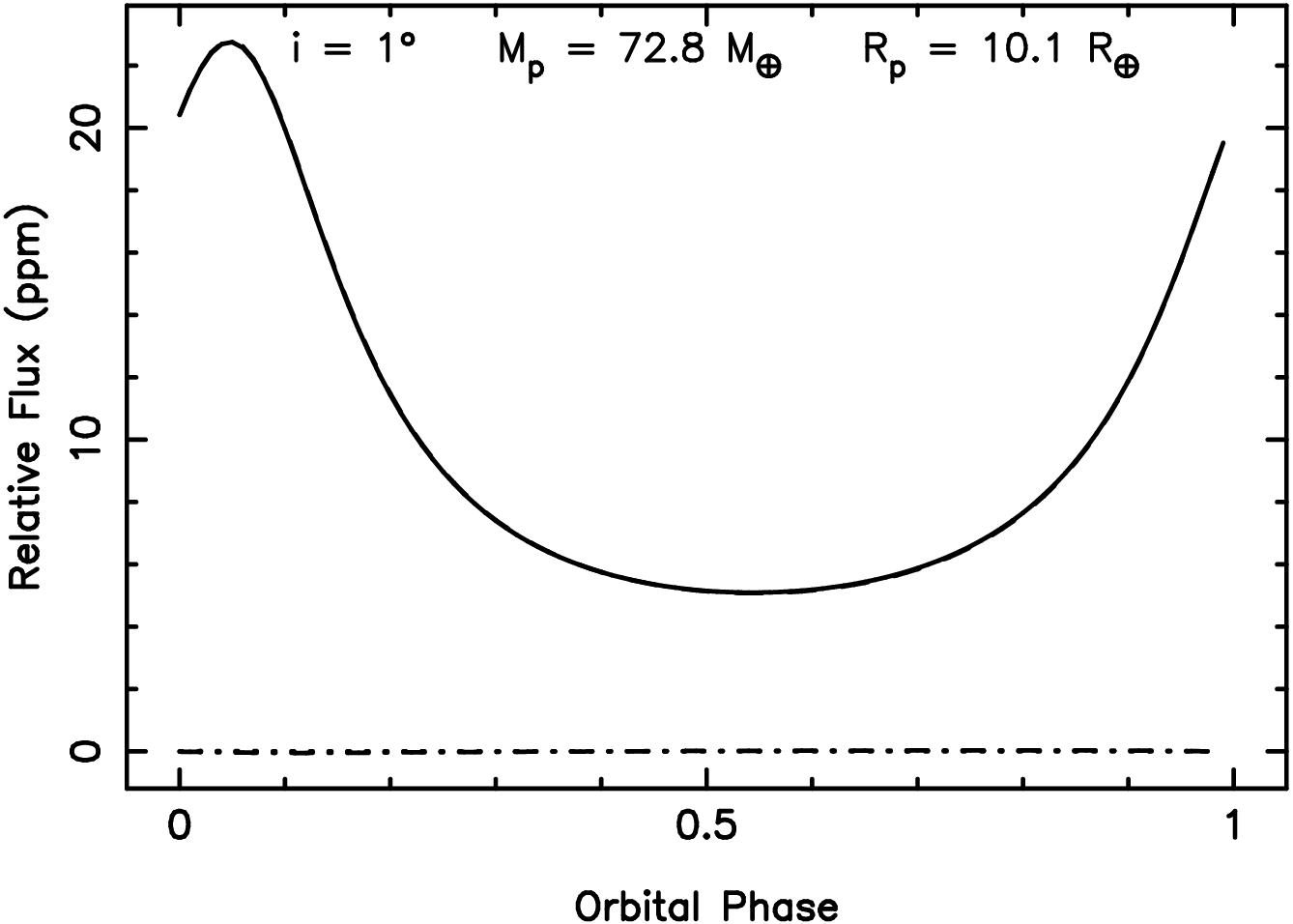}
    \end{tabular}
  \end{center}
  \caption{Flux ratio of Proxima Centauri b to the host star, assuming
    inclinations of $i = 90\degr$ (top-left), $i = 60\degr$
    (top-right), $i = 30\degr$ (bottom-left), and $i = 1\degr$
    (bottom-right). The flux is represented as parts per million
    (ppm). Shown are the contributions of reflected light (dashed
    line), Doppler boosting (dot-dashed line), ellipsoidal variations
    (dotted line), and the combination of all three (solid line).}
  \label{reffig}
\end{figure*}

\section{Reflected Light and Phase Variations}
\label{ref}

Since observations have not currently shown that the Proxima Centauri
planet transits the host star \citep{ang16,dav16,kip16}, detailed
characterization of the atmosphere will likely rely largely upon
reflected/scattered light. The dependence of photometric phase
variations due to reflected light on planetary radii and albedo is
well known \citep{sea00,sud05}, and has also been shown to depend on
orbital eccentricity \citep{kan10}. \citet{kan11a} further
demonstrated how phase variations depend on orbital inclination,
providing a possible mechanism to distinguish between different
classes of orbital objects \citep{kan12a}. \citet{lov16} have
calculated phase amplitudes and contrast ratios for direct detection
of the planet at quadrature points with SPHERE/ESPRESSO. Here we
provide phase variation calculations as a function of orbital phase
and inclination.

\begin{figure*}
  \begin{center}
    \begin{tabular}{cc}
      \includegraphics[angle=270,width=8.5cm]{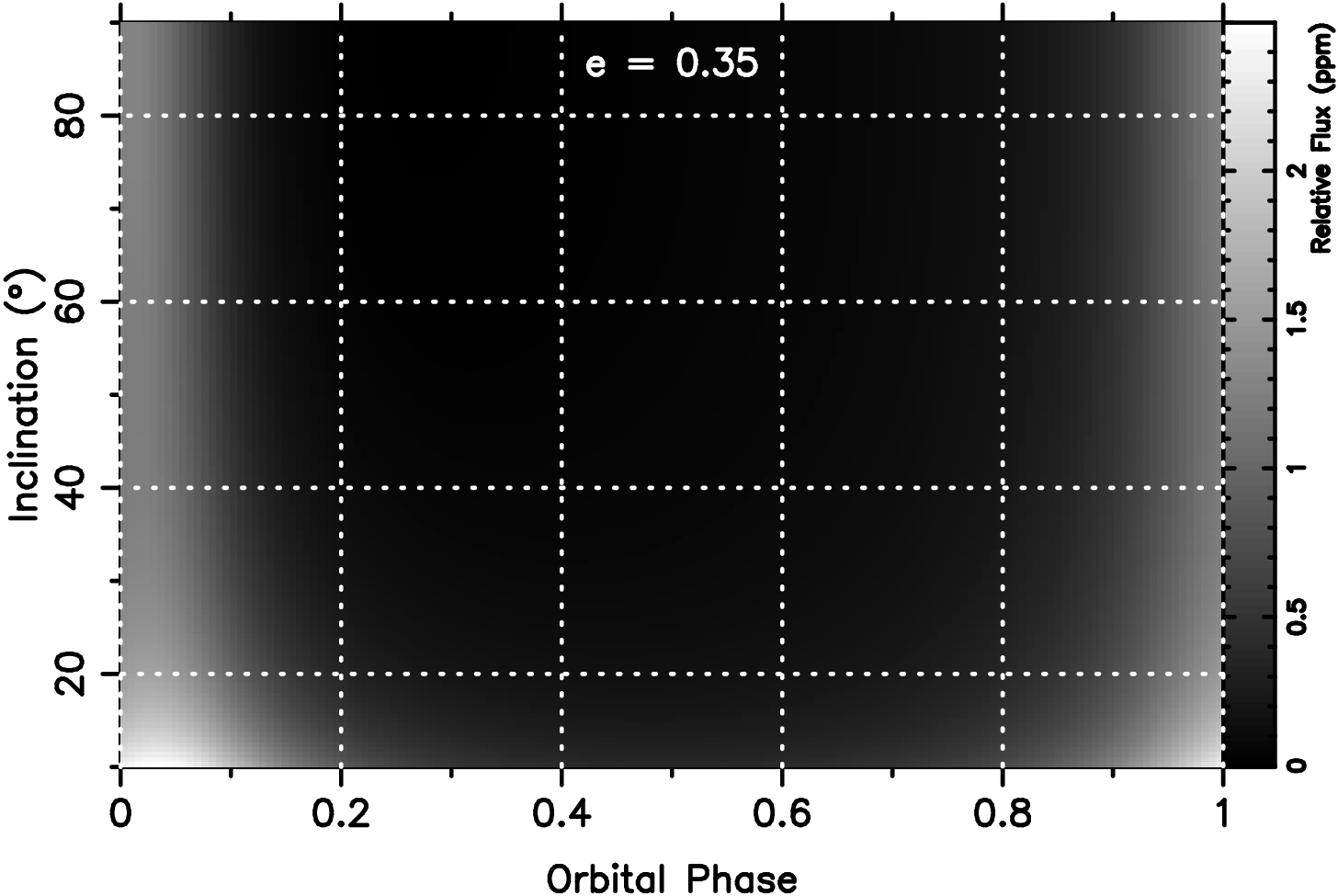} &
      \includegraphics[angle=270,width=8.5cm]{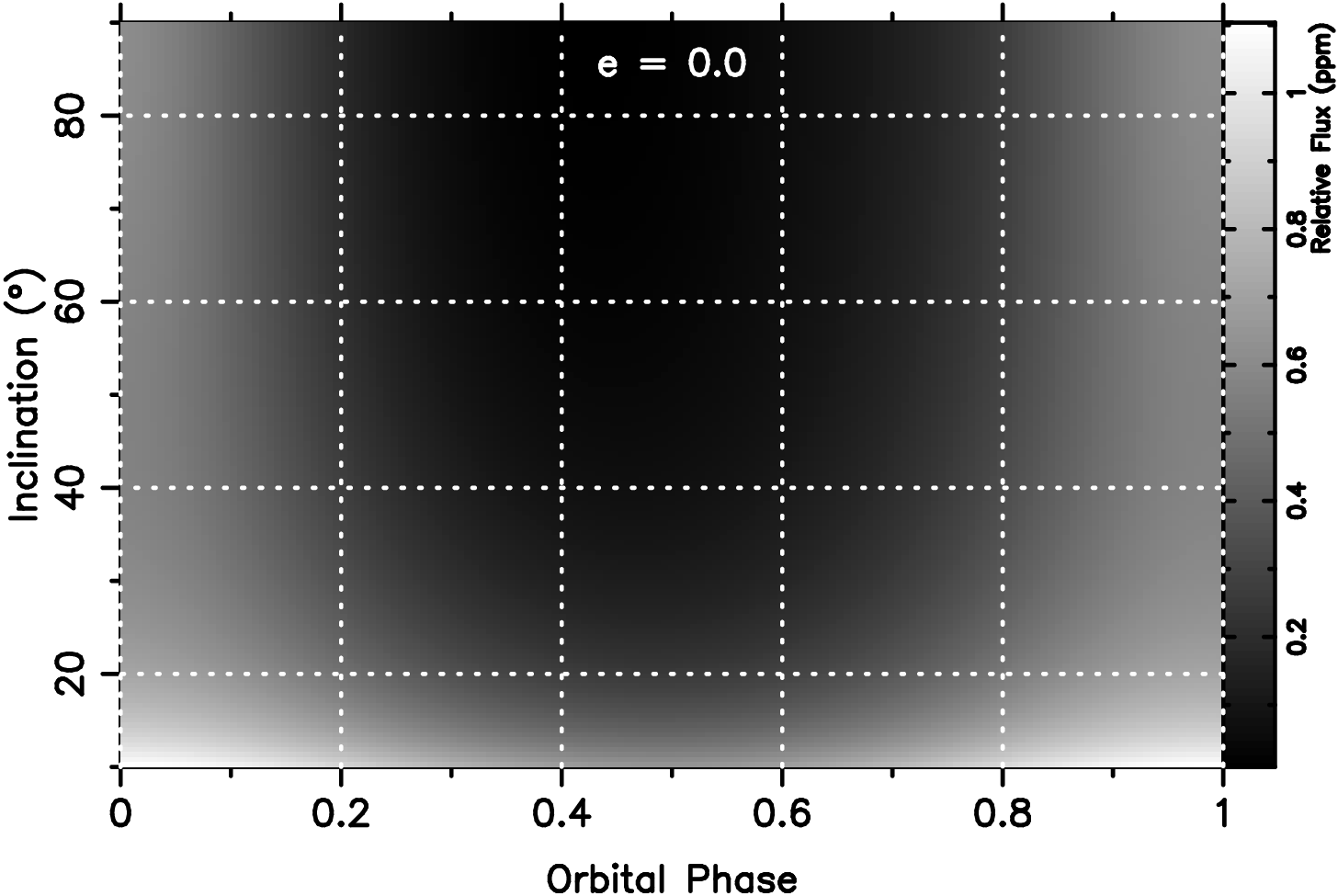}
    \end{tabular}
  \end{center}
  \caption{An intensity map for the flux ratio of Proxima Centauri b
    to the host star, as a function of orbital inclination and orbital
    phase for the eccentric case ($e = 0.35$) and the circular case
    ($e = 0.0$). The flux ratio includes the effects of reflected
    light, Doppler boosting, and ellipsoidal variations.}
  \label{intensitymap}
\end{figure*}

For reflected light at wavelength $\lambda$ and phase angle $\alpha$,
the flux ratio of a planet with radius $R_p$ to the host star is given
by
\begin{equation}
  \epsilon(\alpha,\lambda) \equiv
  \frac{f_p(\alpha,\lambda)}{f_\star(\lambda)}
  = A_g(\lambda) g(\alpha,\lambda) \frac{R_p^2}{r^2}
  \label{fluxratio}
\end{equation}
where $A_g(\lambda)$ is the geometric albedo, $g(\alpha,\lambda)$ is
the phase function, and $r$ is the star--planet separation. The value
of $r$ depends upon the Keplerian orbital elements as follows
\begin{equation}
  r = \frac{a (1 - e^2)}{1 + e \cos f}
  \label{separation}
\end{equation}
where $f$ is the true anomaly. The $R_p^2/r^2$ component of
Equation~\ref{fluxratio} can thus become dominant for highly eccentric
orbits. The phase angle, defined to be zero when the planet is at
superior conjunction, is given by
\begin{equation}
  \cos \alpha = - \sin (\omega + f)
  \label{phaseangle}
\end{equation}
For the phase function $g(\alpha,\lambda)$, we adopt the empirically
derived version of \citet{hil92}, based upon observations of Jupiter
and Venus. This approach uses a visual magnitude correction of the
form
\begin{equation}
  \Delta m (\alpha) = 0.09 (\alpha/100\degr) + 2.39
  (\alpha/100\degr)^2 -0.65 (\alpha/100\degr)^3
  \label{magcorr}
\end{equation}
and the phase function is then given by
\begin{equation}
  g(\alpha) = 10^{-0.4 \Delta m (\alpha)}
  \label{phase}
\end{equation}
For the geometric albedo $A_g(\lambda)$, there are various values that
could be adopted, such as the star--planet separation dependent values
of \citet{kan10}. For the purposes of this study, we adopt a value of
0.5 which is midway between Earth ($A_g = 0.367$) and Venus ($A_g =
0.67$).

Shown in Figure~\ref{reffig} are the predicted changes in relative
flux for Proxima Centauri b over one complete orbit, starting at a
phase angle of $\alpha = 0\degr$. For completeness, we include the
effects of Doppler boosting \citep{loe03,fai11} and ellipsoidal
variations \citep{mor93,zuc07}. The contributions to the total flux
variations (solid line) shown in Figure~\ref{reffig} thus include the
contributions from reflected light (dashed line), Doppler boosting
(dot-dashed line), and ellipsoidal variations (dotted line). As
expected, the Doppler boosting and ellipsoidal variation components
have negligible contributions since they depends largely upon the
planetary mass. These calculations are performed for four different
inclinations, ranging from edge-on ($i = 90\degr$) to face-on ($i =
1\degr$).

The primary change that occurs for the different inclinations is that
the increase in planetary mass leads to an increase in radius, thus
leading to an increase in flux ratio between the planet and star. To
estimate the change in radius, we use the mass-radius relationship
derived by \citet{kan12b}. Of particular interest is that, despite the
loss of phase variations, the reflected light component dominates the
total relative flux variations for face-on orbits due to the
combination of large radius and orbital
eccentricity. Figure~\ref{intensitymap} represents the orbital
inclination dependence of the flux ratio profile as an intensity map,
where the intensity scale is shown on the right of the figure. The
peak flux ratio increases dramatically for orbital inclinations below
$\sim$$10\degr$ due to the rise in planetary radius. The left and
right panels of Figure~\ref{intensitymap} demonstrate the dramatic
change in flux ratio as a function of orbital phase caused by the
orbital eccentricity of the planet.

Thermal phase curves provide an additional avenue towards detection of
the planet, depending on atmospheric composition and dynamics
\cite{sel11,mau12}. For the planet--star contrast ratio at infrared
wavelengths, we calculate their emissions assuming blackbody radiation
and that the planetary atmosphere has 100\% heat redistribution
efficiency \citep{kan11b}. The planetary equilibrium temperature is
then given by
\begin{equation}
  T_p = \left( \frac{L_\star (1-A)}{16 \pi \sigma r^2}
  \right)^\frac{1}{4}
\end{equation}
where $L_\star$ is the stellar luminosity and $A$ is the planetary
spherical (Bond) albedo. The observed contrast ratio at frequency
$\nu$ is given by
\begin{equation}
  \frac{F_p}{F_\star} = \frac
       {(\exp{(h \nu / k T_\mathrm{eff}) - 1)} R_p^2}
       {(\exp{(h \nu / k T_p) - 1)} R_\star^2}
  \label{fluxratio}
\end{equation}
where $T_\mathrm{eff}$ is the stellar effective temperature. As for
the phase variation calculations above, we assume a Bond albedo of $A
= 0.5$. We calculate contrast ratios for the original {\it Spitzer}
passbands of 3.6, 4.5, 8.0, and 24.0~$\mu$m and for the four
inclinations shown in Figure~\ref{reffig}. These passbands are
considered to be representative of the passbands that will be
available at future facilities, such as the 2.4--5.0~$\mu$m wavelength
range of NIRCam on the James Webb Space Telescope (JWST). The
calculated contrast ratios are provided in Table~\ref{contrast}. It is
clear from these numbers that the infrared flux of the planet will be
readily detectable for inclinations less than $30\degr$.

\begin{deluxetable}{ccccc}
  \tablecolumns{5}
  \tablewidth{0pc}
  \tablecaption{\label{contrast} IR Contrast ratios}
  \tablehead{
    \colhead{Inclination} &
    \multicolumn{4}{c}{IR Contrast Ratio (ppm)} \\
    \colhead{($\degr$)} &
    \colhead{3.6~$\mu$m} &
    \colhead{4.5~$\mu$m} &
    \colhead{8.0~$\mu$m} &
    \colhead{24.0~$\mu$m}
  }
  \startdata
  90 & 0.07 &  0.54 &   17.2 &   219.3 \\
  60 & 0.08 &  0.63 &   20.0 &   254.3 \\
  30 & 0.14 &  1.10 &   35.1 &   446.9 \\
  1  & 4.31 & 34.44 & 1101.5 & 14006.8
  \enddata
\end{deluxetable}


\section{Habitable Zone and Orbital Stability}
\label{stab}

\begin{figure}
  \includegraphics[angle=270,width=8.5cm]{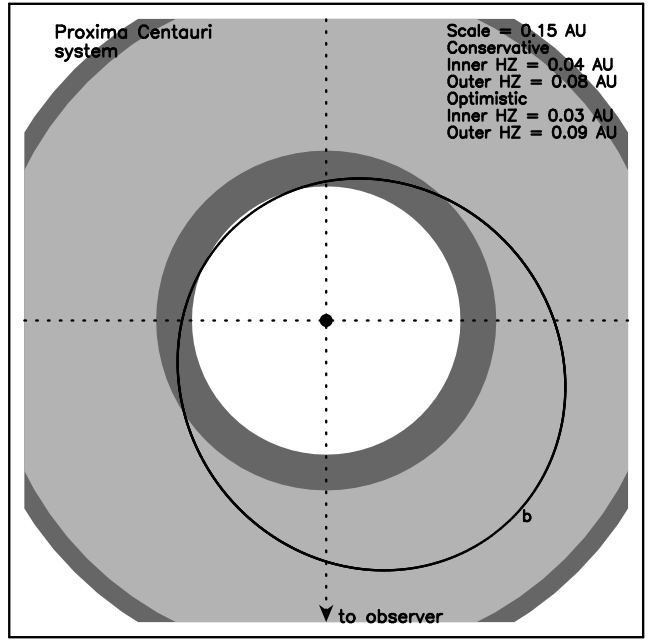}
  \caption{A top-down view of the Proxima Centauri system showing the
    extent of the HZ and orbits of the planets calculated using the
    stellar and planetary parameters from \citet{ang16}. The physical
    scale depicted is 0.15~AU on a side. The conservative HZ is shown
    as light-gray and optimistic extension to the HZ is shown as
    dark-gray.}
  \label{hz}
\end{figure}

A further dependence of the mass of the known planet is the dynamical
stability of additional terrestrial planets in or near the HZ of
Proxima Centauri. To calculate the HZ, we use the stellar parameters
of \citet{ang16} and the methodology of \citet{kop13,kop14}. This
results in estimates for the ``conservative'' and ``optimistic'' HZ
boundaries, the definitions of which depends upon assumptions
regarding the prevalence of liquid water on the surfaces of Venus and
Mars throughout their histories. For the conservative HZ, we calculate
inner and outer boundaries of 0.041 and 0.081~AU respectively. For the
optimistic HZ, we calculate inner and outer boundaries of 0.032 and
0.086~AU respectively. The extent of the HZ and the orbit of the known
planet are depicted in the top-down view of the Proxima Centauri
system shown in Figure~\ref{hz}. The conservative HZ is shown as
light-gray and the optimistic extension to the HZ is shown as
dark-gray. For the eccentric model of the orbit, the planet spends
93\% of the orbital period within the HZ including the optimistic
region.

To test the orbital stability scenarios, we utilize the Mercury
Integrator Package, described in detail by \citet{cha99}. Mercury
performs N-body integrations that are configured with user-supplied
parameters that define the properties and starting conditions for the
system. The specific integrator used was a hybrid
symplectic/Bulirsch-Stoer integrator with a Jacobi coordinate system
since that tends to provide greater accuracy for multi-planet systems
\citep{wis91,wis06}. We use a stability criterion that requires both
planets to remain in the system for the duration of the simulation. If
any of the planets are lost from the system, either by collision with
the host star or ejection from the system, then the system is regarded
as being unstable.

We conducted a series of simulations that place an Earth-mass planet
as a test particle at a range of semi-major axes, from 0.02 to 0.1~AU
in steps of 0.005~AU. Such an orbital range fully encompasses both the
orbit of the known planet and the HZ of the system. We assumed a
circular orbit for the additional planet and used a time resolution of
0.1~days to ensure that the minimum timestep recommendation of
\citet{dun98} ($1/20$ of the shortest system orbital period) was met
at all times. The known planet was assumed to have the maximum allowed
eccentricity of $e = 0.35$. The simulations were conducted for three
different inclination scenarios of $90\degr$, $30\degr$, and
$10\degr$. These inclinations imply a mass for the known planet of
1.27, 2.54, and 7.31~$M_\oplus$ respectively.

\begin{figure*}
  \begin{center}
    \includegraphics[angle=270,width=15.5cm]{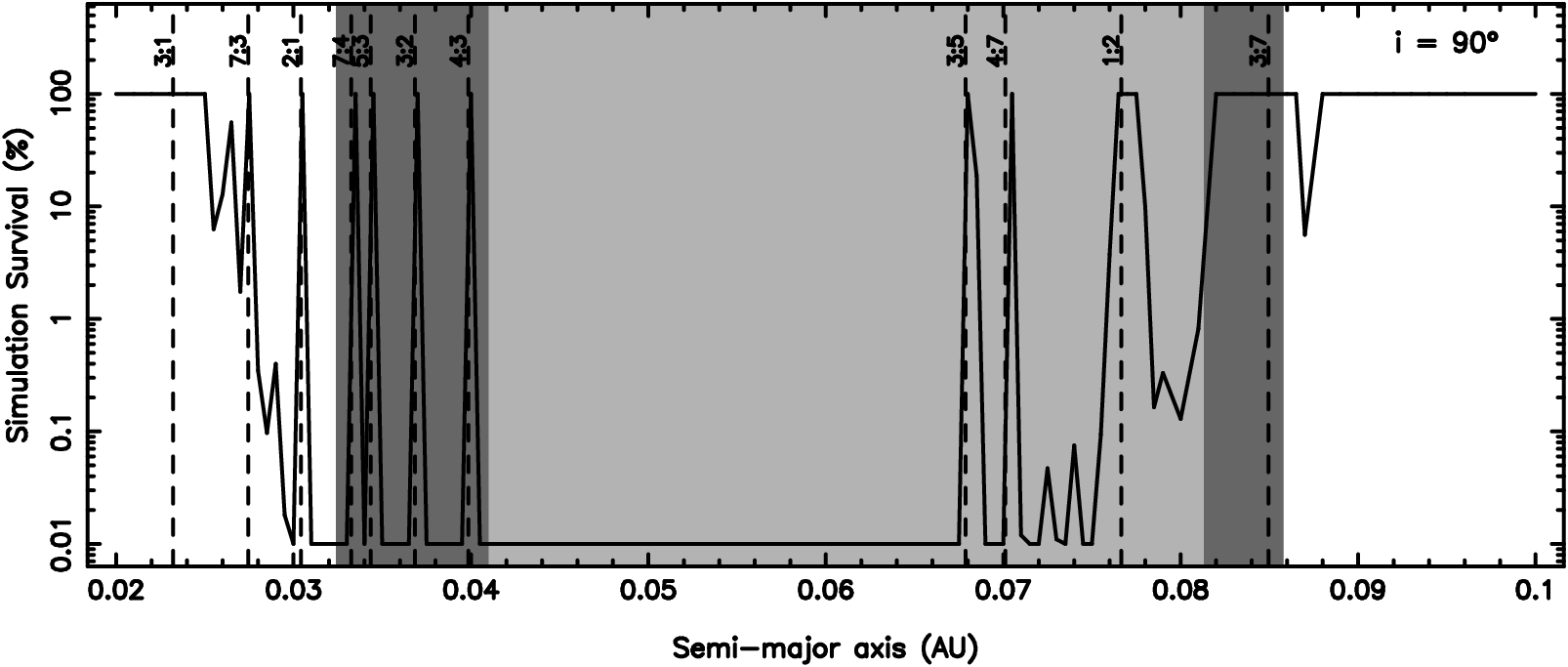} \\
    \includegraphics[angle=270,width=15.5cm]{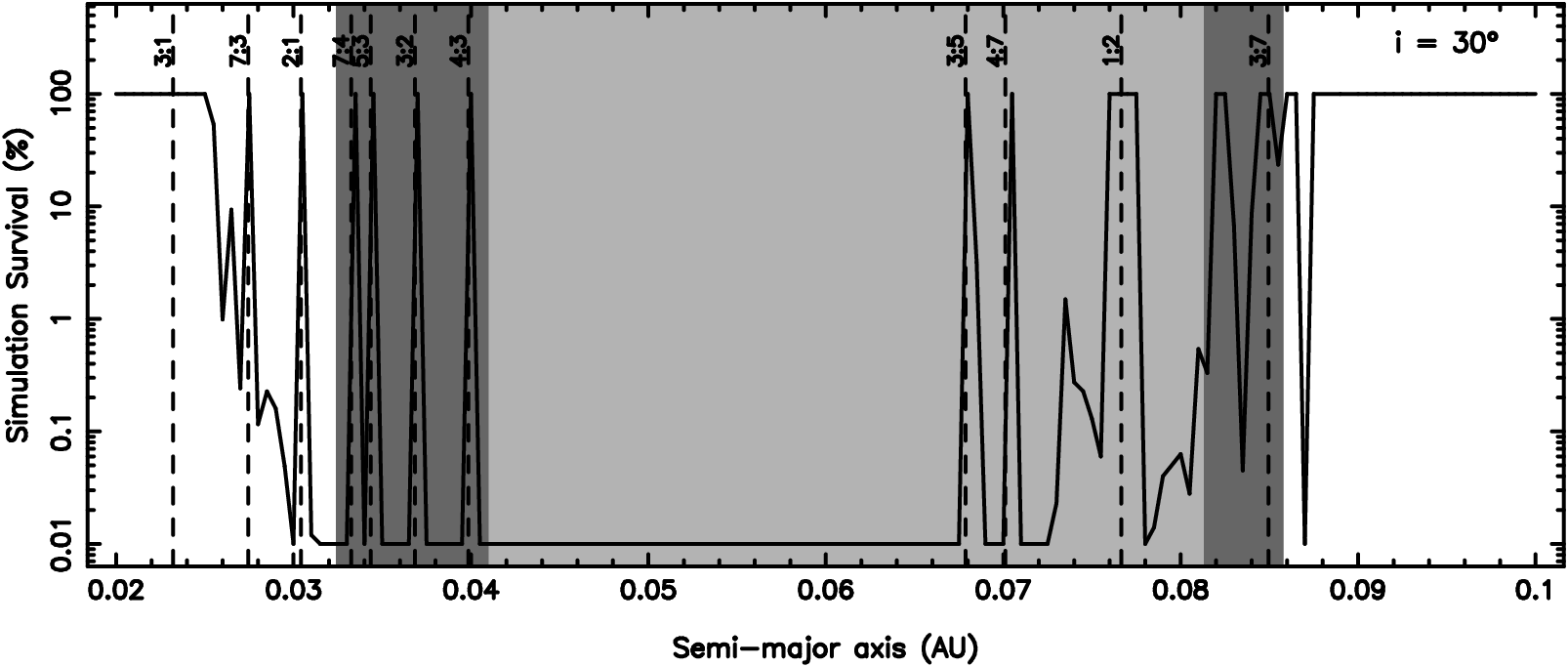} \\
    \includegraphics[angle=270,width=15.5cm]{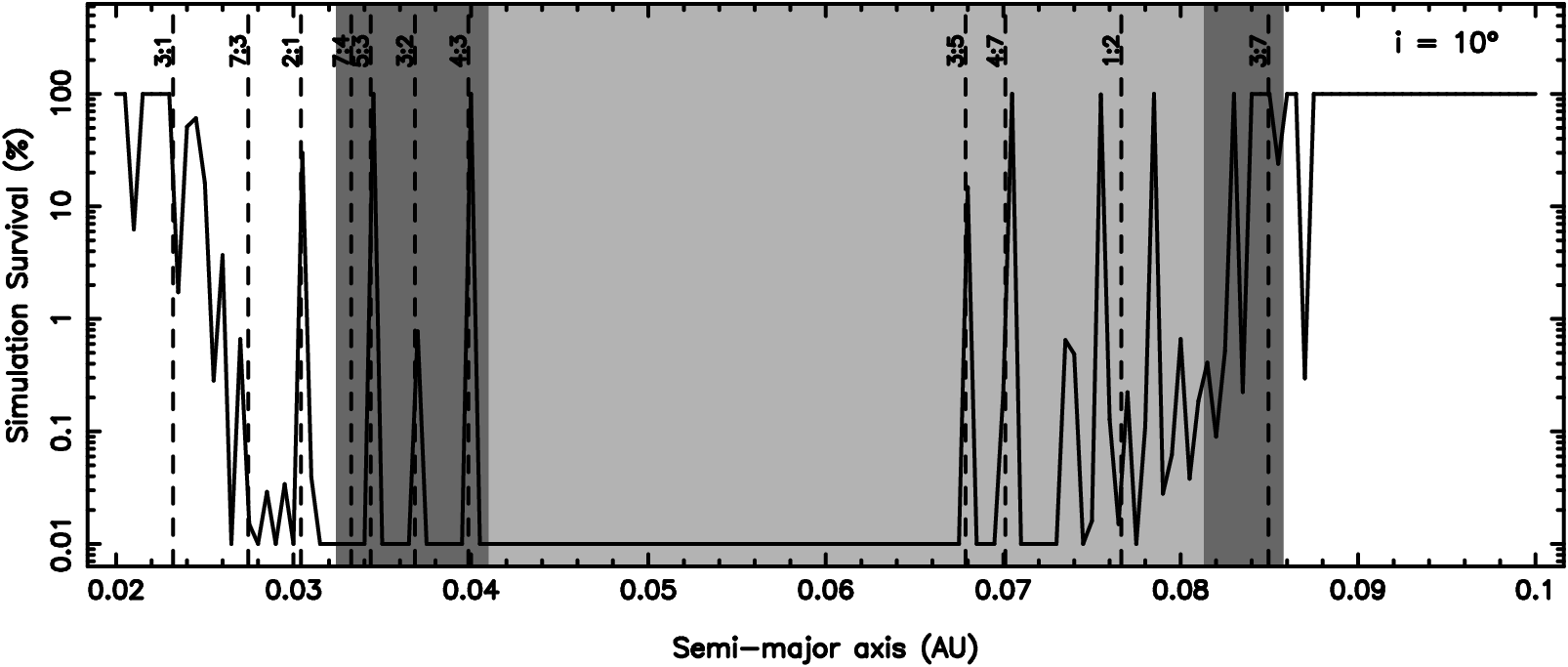}
  \end{center}
  \caption{The orbital stability of a hypothetical Earth-mass planet
    as a function of semi-major axis in the Proxima Centauri
    system. The individual panels show the results for simulations
    that assume the known planet has an inclination of $90\degr$
    (top), $30\degr$ (middle), and $10\degr$ (bottom). These
    inclinations affect the true mass of the known planet and thus the
    overall stability of the system. The orbital stability on the
    vertical axis is expressed as the percentage simulation survival
    for each semi-major axis where the position of the planet was
    tested. The system is assumed to be coplanar and the orbit of the
    Earth-mass planet is assumed to be circular. The light-gray and
    dark-gray regions represent the conservative and optimistic HZ
    regions, as per Figure~\ref{hz}. The vertical dashed lines
    indicate the locations of the mean-motion orbital resonances with
    the known planet.}
  \label{stabfig}
\end{figure*}

The outcome of the orbital stability simulations are shown in
Figure~\ref{stabfig}, where the separate panels show the results for
the $i = 90\degr$ (top), $i = 30\degr$ (middle), and $i = 10\degr$
(bottom) scenarios. As for Figure~\ref{hz}, the conservative HZ is
shown as light-gray and the optimistic extension to the HZ is shown as
dark-gray. For $i = 90\degr$, the presence of the known planet
excludes other planets within the HZ with the exception of the
locations of mean-motion resonance (MMR), shown in
Figure~\ref{stabfig} as vertical dashed lines. Outside of the HZ, the
dynamical viability of additional planets rises dramatically. There is
very little difference between the $90\degr$ and $30\degr$ cases,
since the Hill radius has a mass dependence of $M_p^{(1/3)}$ but
scales linearly with $a$. Thus, the close proximity of the planet(s)
to the host star dominates the orbital dynamics and subsequent
stability, as observed for compact systems, such as those found by
Kepler \citep{ray09}. Note however that the 3:7 MMR narrows
significantly between the $90\degr$ and $30\degr$ cases, reducing its
viability as an orbital location for another planet.

The third scenario that we investigated was the case for $i =
10\degr$, where the mass of the known planet would be
$\sim$7.31~$M_\oplus$. The result of this simulation is shown in the
bottom panel of Figure~\ref{stabfig}. The main effect of the planetary
mass increase is to further erode the significance of the MMR
locations, rendering them largely unstable. The exception to this is
the emergence of two stability locations on either side of the 1:2
MMR. An additional effect of the interaction between the two planets
is the exchange of angular momentum, resulting in oscillating
eccentricities of the known planet (for example, see
\citet{kan14}). Tidal effects will also undoubtedly play a role in the
potential habitability of the known planet \citep{bar09} as well as
circularizing the orbit \citep{bar16}. If the eccentricity of the
known planet is close to the maximum of 0.35 found by \citet{ang16},
the planet may spend extended periods in a close to circular orbit
within the HZ as the other planet increases in eccentricity.

Finally, we tested the case of a circular ($e = 0.0$) edge-on ($i =
90\degr$) orbit for Proxima Centauri b with the addition of the
hypothetical Earth-mass planet described above. In this case, the
orbital stability of the system is preserved for all semi-major axes
of the additional planet outside of the range
0.044--0.053~AU. Comparison of this range with the instability regions
depicted in Figure~\ref{stabfig} shows that a circular orbit for
planet b allows there to be significantly more locations where another
low-mass planet could be harbored by the system in a stable orbit than
for the eccentric case. This result emphasizes the dependence on
orbital eccentricity and the need to fully understand the Keplerian
nature of the planet b orbit.


\section{Observational Prospects}
\label{obs}


\subsection{Near-term Characterization}

Clearly, further observations are required to resolve the inclination
ambiguity described in this work and determine whether this world
could be habitable. In Section~\ref{ast}, we demonstrated that an
astrometric signal for this planet is within the sensitivity regime of
{\it Gaia}, if the planet is a gas giant. Thus we may have
confirmation of the planet's terrestrial nature within the next
several years of {\it Gaia} data releases. Meanwhile, JWST may offer a
near-term prospect for constraining the planet's size and atmospheric
thermal properties for low inclination orbits (see
Section~\ref{ref}). \citet{kri16} further find that with complete
phase coverage, JWST could detect variations in thermal emission with
a precision sufficient to distinguish between bare rock and a planet
with partially (35\%) redistributed heat due to the presence of an
atmosphere and/or ocean.


\subsection{Direct Imaging from the Ground}

Ultimately, the greatest promise for assessing the habitability (or
inhabitance) of Proxima Centauri b lies with directly imaging the
planet and determining atmospheric composition via spectral
characterization. In Section~\ref{ast}, we calculated a maximum
angular separation between Proxima Centauri and planet b of
$\sim$65~mas, depending on inclination and eccentricity. In
Section~\ref{ref}, we calculate a planet-to-star flux ratio in
reflected starlight of $10^{-5}$ to $10^{-6}$, with flux variations at
levels of $10^{-6}$ or more over the course of an orbit, due to phase
changes.

To date, this star--planet separation and flux contrast is
substantially smaller than what has been achieved with direct imaging
even for young, self-luminous planets (e.g., \citet{mac15}). As
mentioned in Section~\ref{ref}, \citet{lov16} have suggested the
possibility of directly imaging Proxima Centauri b by combining the
SPHERE coronagraph with the ESPRESSO spectrograph at the 8m ESO
VLT. They calculate a detection time of 20--40 nights of telescope
time, and possible atmospheric O$_2$ detection in 60 nights. In the
more distant future, extremely large ($\sim$40m class) ground-based
observatories could offer the capability required to spectrally
characterize this planet in the optical and near-IR. However,
\citet{mea16} point out that shorter wavelength coverage than is
currently planned for the E-ELT and GMT adaptive optics capabilities
would be desirable.


\subsection{Direct Imaging from Space}

Space-based coronograph and/or starshade missions may provide the
greatest capability in revealing the nature of the Proxima Centauri
planet, through spectral characterization in the UV through
near-IR. The Wide-Field Infrared Survey Telescope (WFIRST) will be the
first technology demonstration of wavefront controlled space-based
exoplanet imaging coronagraphs, and it is scheduled for launch in
$\sim$2025. The highest priority WFIRST targets will be chosen from
the brightest known RV planets. The most ambitious of the WFIRST
coronagraph designs was the phase-induced amplitude apodization
complex mask coronagraph (PIAACMC). PIAACMC's assumed inner working
angle and contrast floor (40~mas and $\sim$$3.4 \times 10^{-10}$;
\citet{tra16}) would have been sufficient to detect the planet at
locations near the maximum angular separation found in this
paper. However, the PIAACMC was designated as a ``backup'' instrument
due to outstanding technical challenges, and it is unlikely to advance
beyond Phase A. The baseline WFIRST Hybrid Lyot and Shaped Pupil
coronagraphs will achieve inner working angles of 120--150~mas in the
shortest wavelength band at 465~nm, but this performance is not
sufficient to image Proxima Centauri b.

The WFIRST baseline mission also includes starshade readiness, which
leaves open the possibility that a separately launched starshade could
rendezvous with WFIRST later in the mission. \citet{sea15} found that
such a rendezvous mission could achieve an inner working angle of
$\sim$70~mas in the bluest bandpass (425--600 nm), and contrast limit
of better than $10^{-10}$. This is approaching to the necessary inner
working angle required for detecting Proxima Centauri b, and perhaps
the design could be further optimized for this target. However, the
large size of the WFIRST point spread function ($\sim$50~mas) may
nevertheless lead to prohibitively long exposure times even with the
high throughput provided by a starshade.

Therefore, we conclude that the spectral characterization of Proxima
Centauri b in the UV through near-IR may have to wait for larger
space-based concepts like the 4--6.5m ``Hab-Ex'' Habitable Exoplanets
Imaging Mission\footnote{\tt http://www.jpl.nasa.gov/habex/} or the
8--12m LUVOIR\footnote{\tt
  http://cor.gsfc.nasa.gov/studies/luvoir.php} currently under
study. It is worth noting that a prerequisite for scheduling
observations times is the refinement of the planetary orbit to produce
an accurate ephemeris \citep{kan09}.


\section{Conclusions}
\label{conclusions}

Proxima Centauri b is the closest exoplanet to our planetary system,
and thus provides interesting prospects for further
characterization. The terrestrial nature of the planet is quite
likely, and is calculated by us to be $\sim$84\% (see
Section~\ref{effect}). Furthermore, previous studies have shown that
giant planets in short-period orbits around M dwarfs are relatively
rare \citep{bon13,tuo14,dre15a}. However, the ambiguity regarding the
mass of the planet due to the unknown orbital inclination will greatly
influence the outcome of further investigations. The mass of the
planet will have a profound impact on such other properties as radius,
atmospheric scale height, composition, and albedo. Each of these
properties, in turn, will determine the detectability of the planet
via alternative methods.

In this work, we have quantified the astrometric signature and angular
separation of the planet as a function of inclination. Our astrometry
calculations show the region of inclinations and astrometric
amplitudes where the planet can be considered to have crossed from the
terrestrial into the gas giant regime. Although a close to face-on
inclination would entail the planet not being of terrestrial mass, the
angular separation calculations show that this scenario produces the
largest angular separation and the least constraints on direct imaging
observations.

Our calculations of the expected phase variations as a function of
inclination show that the face-on scenarios produce the largest
amplitude. However, face-on phase amplitudes are being driven by a
large reflecting area (radius) and a time-dependent star--planet
separation, and so depends highly upon the eccentricity of the
orbit. For inclinations where $i > 30\degr$, there is very little
difference in the overall shape and amplitude of the phase variations,
including the components of reflected light, Doppler boosting, and
ellipsoidal variations. The change in inclination has a dramatic
effect on the predicted contrast ratio at infrared wavelengths, such
as a contrast ratio of $\sim$1\% at 24~$\mu$m for an inclination of $i
= 1\degr$.

We calculated the extent of the optimistic and conservative HZ for
Proxima Centauri and conducted exhaustive dynamical simulations to
determine the viability of other terrestrial planets within the HZ
region. Our simulations demonstrate that the presence of the known
planet with an eccentric orbit excludes the possibility of another
terrestrial planet throughout most of the HZ with the exception of MMR
locations. Reducing the inclination to $i = 10\degr$ further compounds
the instability within the HZ regions.

The overall results contained within this work are meant to serve as a
guide for future observations intended to characterize the planet,
particularly those that may detect reflected light or direct emission
from the planet. Such observations may include proposed coronographs
or similar instruments for future space-based imaging missions. Other
techniques beyond those discussed here, such as microlensing
\citep{sah14}, may also benefit from our quantification of the
observable signatures. Given that the Proxima Centauri planet is not
only the closest exoplanet, but the nearest planet in the HZ of its
host star, the potential rewards for further studies are highly
warranted.


\section*{Acknowledgements}

The authors would like to thank the anonymous referee, whose comments
improved the quality of the paper. Thanks are also due to Guillem
Anglada-Escud\'e and Franck Selsis for their useful feedback on the
manuscript. This research has made use of the following archives: the
Exoplanet Orbit Database and the Exoplanet Data Explorer at
exoplanets.org, the Habitable Zone Gallery at hzgallery.org, and the
NASA Exoplanet Archive, which is operated by the California Institute
of Technology, under contract with the National Aeronautics and Space
Administration under the Exoplanet Exploration Program. The results
reported herein benefited from collaborations and/or information
exchange within NASA's Nexus for Exoplanet System Science (NExSS)
research coordination network sponsored by NASA's Science Mission
Directorate.


\end{document}